\documentstyle[12pt]{article}

\def\br{\begin{thebibliography}{abc}}
\def\er{\end{thebibliography}}

\begin{document}

\baselineskip=24pt

\thispagestyle{empty}
\setcounter{page}{0}

\begin{flushright} SU-4240-566 \\
 February 1994
\end{flushright}
{\large
\centerline{\bf GRAVITY AND ELECTROMAGNETISM IN}
\centerline{~~}
\centerline{\bf NONCOMMUTATIVE GEOMETRY}
 }
\vskip 1.5cm
\centerline{ {\bf Giovanni Landi}
\footnote{\noindent On leave from Dipartimento di
Matematica, Universit\`a di Trieste, P.le Europa 1, I-34100, Trieste,
Italy. Also INFN, Sezione di Napoli, Napoli, Italy. \\
Fellow of the Italian National Council of Research (CNR) under Grant
No. 203.01.60.},
{\bf Nguyen Ai Viet}
\footnote{\noindent
On leave from Centre of Theoretical Physics, P.O.Box 429
Bo Ho 10000, Hanoi, Vietnam.} \hskip 3mm
{\bf and Kameshwar C.Wali}}
\vskip .5cm
\centerline{
{\normalsize  Physics Department, Syracuse University, Syracuse, New York
13244-1130}}
\vskip 1.5cm
\begin{abstract}
We present a unified description of gravity and electromagnetism in
the framework of a $Z_2$ noncommutative differential calculus.
It  can be considered as a ``discrete version" of
Kaluza-Klein theory, where the fifth continuous dimension is replaced by two
discrete points. We derive an action which coincides with the
dimensionally reduced one of the ordinary Kaluza-Klein theory.

\end{abstract}
\hspace{2cm}

\newpage

Recently, Connes \cite{Co1,Co2} has proposed a new construction of
noncommutative geometry
where the basic objects are
an algebra ${\cal A}$ (possibly noncommutative), a Hilbert space
on which the algebra acts, and a ``Dirac operator" on the Hilbert space. The
algebra generalizes the idea of a manifold and the operator prescribes
the rule for differentiation and provides the metric structure.

Within the general framework of Connes,
different realizations of the geometry
with different algebras and different Dirac operators have been proposed
leading to various applications in particle physics \cite{MANY}. One such
version due
to Connes and Lott  \cite{CL}, is based on regarding the space-time to
be a direct product of a Riemannian manifold and a discrete ``two-point" space.
A gauge field on such a generalized space-time consists of the usual gauge
field in the Riemannian manifold and a Higgs field as a part of the connection
in the discrete internal space. Thus, it lends itself to a unified geometric
description of the bosonic fields including both the ordinary gauge fields and
Higgs fields. And what is remarkable is that the action constructed on such a
generalized space contains a spontaneous symmetry-breaking Higgs
potential.

An alternate way of picturing this generalized space-time is along the lines of
Kaluza-Klein theories  \cite{KK},
in which the continuous extra fifth dimension describing
the internal space is replaced by a finite number of discrete points.
The advantage of this approach over the ordinary Kaluza-Klein theory
is that there is no truncation of any physical modes, while in the
latter an infinite number of massive modes is truncated.
Since this is a new
concept of space-time, it is natural to investigate its implication on gravity
within the framework of noncommutative geometry. A first step in this direction
was taken by Chamseddine, Felder and Fr\"ohlich, who employed a vielbein and a
connection in generalized Cartan structure equations. This has led to
Einstein's gravity along with
a massless scalar field coupled to gravity and a
cosmological constant \cite{CFF}. The Wodzicki residue has also been
used to obtain gravity with a cosmological constant  \cite{KW}.
Since the analysis of these authors can be extended to a
generalized space consisting of an arbitrary number of discrete points, the
question arises regarding how the vector fields in the Kaluza-Klein theory
disappear when we approximate the compact internal space by a finite number of
discrete points. Does discretization eliminates the vector modes altogether?

This note addresses the above question. We show that in the simplest version of
a generalized space-time with two discrete points, one can formulate a
noncommutative geometry which includes a symmetric tensor, a vector and a
scalar fields. In other words, we can construct a Kaluza-Klein type unified
Einstein-Maxwell theory. In fact, in  \cite{KW} a
vector field was included in the connection and was shown that it
dropped out completely.
As we shall
see, this does not happen if the vector fields are introduced in the
vielbein \footnote{After completing the work we received a
preprint by Sitarz \cite{Si}, which
has some similar elements in the mathematical
construction as ours, but the final result is not different from the previous
ones for the same reason.}.

In this brief note, we shall limit the mathematical
details to the necessary minimum and concentrate on the physical aspects of the
model. More details will appear elsewhere  \cite{LVW}.

\bigskip

We consider a noncommutative geometry {\it \`a la} Connes \cite{Co2} built on
a $Z_2$
algebra, $\Omega^0={\cal A}= C^\infty({\cal M})\times Z_2$, where
$C^\infty({\cal M})$ is the algebra of smooth real functions on the
four dimensional manifold
${\cal M}$. That is to say, any element $F \in \Omega^0 $ can be expressed as
\begin{equation}
F(x)=\tilde f_1(x) e + \tilde f_2(x) r~ ,
\end{equation}
where
\begin{equation}
e,r \in Z_2 : e^2=e ,~ er =re=r ,~ r^2=e~.
\end{equation}
We can realize this algebra  ${\cal A}$ by mapping it into the algebra of
$2\times 2$ matrices,
\begin{eqnarray}
\pi(e)=\pmatrix{1&0\cr
                0&1\cr}     &,&  \pi(r)=\pmatrix{1&0\cr
                                                 0&-1\cr}~, \nonumber\\
\pi(F)=\pmatrix{f_1(x)&0\cr
                0&f_2(x)\cr} &=&  \tilde f_1(x)\pmatrix{1&0\cr
                                                        0&1\cr} + \tilde f_2(x)
\pmatrix{1&0\cr
         0&-1\cr}~,
\end{eqnarray}
where $f_1, f_2$ are obvious combinations of $\tilde f_1, \tilde f_2$.

The second crucial ingredient of noncommutative geometry is the Dirac operator.
We will follow the idea of Connes to construct it in parallel with the ordinary
commutative geometry. Let ${\cal E}_N$ ($ N=\mu,5$) be a linearly independent
`tangent basis' which acts on $\Omega^0$ by the commutator,
\begin{equation}
{\cal E}_N(F)= [{\cal E}_N, F].
\end{equation}
It has the following realization
\begin{eqnarray}
&&{\cal E}_\mu = \pmatrix{\partial_\mu &0\cr
                          0 &\partial_\mu\cr} ,~~~\mu = 1, \cdots, 4~,
\nonumber \\
&&{\cal E}_5 = \pmatrix{0& m\cr
-m&0\cr}  ,
\end{eqnarray}
where $m$ is a c-number with the dimension of mass.

It is easy to verify that ${\cal E}_N$ satisfies the Newton-Leibnitz rule,
\begin{equation}
{\cal E}_N(FG) = {\cal E}_N(F) G + F {\cal E}_N(G).
\end{equation}
Hence, we can consider ${\cal E}_N$ as derivations in the $Z_2$-
noncommutative geometry and denote them by $D_N\doteq {\cal E}_N$.
The possibility to enlarge the `tangent basis' by an outer automorphism $
D_5$ enriches the structure of noncommutative geometry. Without it the
geometry is of commutative character.

Working in the Hilbert space of spinors, Connes  \cite{Co2,CL}
chose the `cotangent basis'
$\Gamma^M$ of the extended Dirac matrices as follows
\begin{eqnarray}\label{dirac}
\Gamma^\mu = \pmatrix{\gamma^\mu & 0\cr
                        0& \gamma^\mu\cr} ,~~~
\Gamma^5 = \pmatrix{\gamma^5&0\cr
0&-\gamma^5\cr}.
\end{eqnarray}

Hence, the Dirac operator in the $Z_2$-noncommutative geometry has the
self-adjoint realization
\begin{eqnarray}
\not \!\!D \doteq \Gamma^N D_N \equiv \pmatrix{\not\!\partial &\gamma^5
m\cr
                                           \gamma^5 m
&\not\!\partial\cr}~.
\end{eqnarray}

The choice (\ref{dirac}) of $\Gamma^5$ leads to
\begin{equation}
\{\not \!\! D, \Gamma^5 \} = 0~ .
\end{equation}
Since in this paper we are not working in the Hilbert space of spinors, we need
an alternative realization for the `cotangent basis' $\epsilon^N$. We can
choose
the realization
\begin{eqnarray}
DX^\mu &\equiv & \epsilon^\mu \doteq \pmatrix{dx^\mu &0 \cr
                                     0& dx^\mu \cr} ,~~~ \mu = 1,
\cdots, 4~, \nonumber\\
DX^5 &\equiv & \epsilon^5 \doteq \pmatrix{\theta& 0\cr
                                   0& -\theta \cr},
\end{eqnarray}
where $\theta $ is a Clifford element satisfying
\begin{equation}
\theta^2= 1~~~, ~~~ \theta dx^\mu = -dx^\mu \theta~.
\end{equation}
In our formalism $\theta $ replaces $\gamma^5 $ when we are not
working on the Hilbert space of spinors.

The exterior derivative operator $D$ is given by
\begin{eqnarray}
D \doteq DX^N D_N \equiv \pmatrix{d&\theta m\cr
                                  \theta m & d\cr}~,
\end{eqnarray}
where $d$ denotes the exterior derivative on ${\cal M}$.
The exterior derivative acts on a function $F = (f_1, f_2)
\in \Omega^0$ as follows:
\begin{eqnarray}
DF\doteq DX^N D_NF = \pmatrix{df_1& \theta m(f_2-f_1)\cr
                             \theta m(f_1-f_2) &df_2\cr} ,
\end{eqnarray}
or in the $\Gamma $-realization
\begin{eqnarray}
\not \!\!D F \doteq [\not \!\!D, F] = \pmatrix{\not \!\partial f_1 & \gamma^5
m(f_2-f_1)\cr
\gamma^5 m (f_1-f_2) & \not \!\partial f_2 \cr}.
\end{eqnarray}

A `vector field' $V\in \Lambda^1$ in the $Z_2$-noncommutative geometry can be
defined as follows
\begin{eqnarray}V \doteq V^N D_N
= \pmatrix{v_1^\mu(x) \partial_\mu & m v_1(x) \cr
                                m v_2(x)  &v^\mu_2(x)\partial_\mu\cr},
\end{eqnarray}
where $V^N$ are elements of $\Omega^0$.

A `covector field' or 1-form  $U\in \Omega^1$ is defined as
\begin{eqnarray}
U\doteq DX^N U_N = \pmatrix{dx^\mu u_{1\mu}(x)& \theta u_1(x)\cr
                              \theta u_2(x) & dx^\mu u_{2\mu }\cr },
\end{eqnarray}
where
\begin{eqnarray}\label{FIRULE}
U_5 \doteq \pmatrix{0&1\cr
                   -1&0\cr} U~,
\end{eqnarray}
and $U^\mu, U$ are element of $\Omega^0$.

The definition of the fifth component of the 1-form by the
rule (\ref{FIRULE}) is
implied by the natural requirement that the element $DF$ be a 1-form.

It is possible to construct higher differential forms and a differential
algebra from the following definition of the wedge product
\begin{eqnarray}
DX^\mu \wedge DX^\nu & \doteq & \pmatrix{ dx^\mu \wedge dx^\nu & 0\cr
0& dx^\mu \wedge dx^\nu\cr}  \equiv  -DX^\nu \wedge DX^\mu ,\nonumber\\
DX^5 \wedge DX^\mu & \doteq & \pmatrix{ \theta dx^\mu & 0\cr
 0 & \theta dx^\mu\cr} \equiv -DX^\mu \wedge DX^5 , \nonumber\\
DX^5 \wedge DX^5 &\doteq & 0
\end{eqnarray}
Alternately, we could have postulated $ DX^5 \wedge DX^5 \not= 0$ and
recover
the construction of Coquereaux et al.  \cite{MANY}. Our construction
treats ``$X^5$" as an `even' element on
an equal footing with the space-time coordinates ``$X^\mu$".

A general $p$-form $W_p \in \Omega^p$ is defined as
\begin{equation}
W_p \doteq DX^{N_1}\wedge...\wedge DX^{N_p} W_{N_1 ... N_p}.
\end{equation}
By generalizing the rule (\ref{FIRULE}), we can express the components
$W_{N_1....N_p}$ in the form,

\begin{equation}
W_{N_1....N_p} = \pmatrix{ 0 & 1\cr
                           -1& 0 \cr}^r W_{\nu_1....\nu_{p-r}},
\end{equation}
where r=0,1 ; $\nu_i$ is the i-th index, which is different from 5 and
$W_{\nu_1...\nu_{p-r}}\in \Omega^0$. If the index 5 appears more than
once, the component is zero.

The exterior derivative $DW_p \in \Omega^{p+1}$ of a $p$-form $W_p \in
\Omega^p$ and the wedge product $ W_{1p} \wedge W_{2q} \in
\Omega^{p+q}$ of a $p$-form $W_{1p}\in\Omega^p$ and a q-form
$W_{2q}\in \Omega^q$ are defined to be
\begin{eqnarray}
DW_p & = & DX^M \wedge DX^{N_1}\wedge....\wedge DX^{N_p} D_M W_{N_1...N_p},
\nonumber \\
W_{1p}\wedge W_{2q} & = & DX^{N_1}\wedge ...\wedge DX^{N_p}\wedge DX^{N_{p+1}}
\wedge...\wedge DX^{M_{p+q}}W_{1 N_1...N_p}W_{2 N_{p+1}...N_{p+q}}.~~~
\end{eqnarray}
We have the following essential properties for the exterior
derivative,
\begin{eqnarray}
&& D^2W_p = 0~,~~~~~\forall~ p~, \nonumber\\
&& D(W_p\wedge W_q) = DW_p \wedge W_q + (-1)^p W_p\wedge DW_q~.
\end{eqnarray}
The noncommutative character of our geometry is reflected in the fact that $W_p
\wedge W_q$ and $W_q\wedge W_p$ are not related in general to each other by a
simple factor as in the case of ordinary commutative geometry.

Although in what follows the geometrical objects we construct resemble
those of ordinary geometry, their noncommutative character dictates
strictly the order.

Next we introduce an orthonormal basis of
vielbein $\{E^A\}~ (A=a,5)$. As a direct generalization of vielbein, $E^A$ are
1-forms in the $Z_2$-noncommutative geometry, $ E^A \doteq DX^M E^A_M $,
whose general expression is as follows,
\begin{eqnarray}
E^a &\doteq & \pmatrix{e^a_1& \theta f_1^a\cr
                   \theta f^a_2 & e_2^a\cr} ,~~ a = 1, \cdots, 4~,\nonumber\\
{}~~ \nonumber \\
E^5 &\doteq & \pmatrix{a_1& \theta \phi_1\cr
                   \theta \phi_2 & a_2\cr} ,
\end{eqnarray}
where $e^a_1, e^a_2$ are vielbein on ${\cal M}$,
$a_1, a_2$ are 1-forms on ${\cal M}$ and
$f_1^a, f_2^a, \phi_1, \phi_2 $ are real function on ${\cal M}$.

As in the usual Riemannian geometry, we still have a degree of freedom to
choose the following forms for vielbein without any loss of generality:
\begin{eqnarray}\label{gvielb}
E^a &\doteq &\pmatrix{e^a_1& 0\cr
                   0 & e_2^a\cr},~~ a = 1, \cdots, 4~, \nonumber\\
{}~~ \nonumber \\
E^5 &\doteq & \pmatrix{a_1& \theta \phi_1\cr
                   \theta \phi_2 & a_2\cr} .
\end{eqnarray}

In this paper, we are particularly interested in the self-adjoint vielbein
\begin{eqnarray}\label{svielb}
E^a &\doteq& \pmatrix{e^a& 0\cr
                   0 & e^a\cr} = DX^\mu e^a_\mu \nonumber\\
{}~~ \nonumber \\
E^5 & \doteq & \pmatrix{a& \theta \phi \cr
                   \theta \phi & a\cr} = DX^\mu a_\mu + DX^5 \pmatrix{0 &1\cr
                                -1&0\cr} \phi(x).
\end{eqnarray}
while the general case will be treated elsewhere \cite{LVW}\footnote{ In the
Eq.(\ref{svielb}), by setting $a=0$ we obtain the vielbein used by Chamseddine
et al. \cite{CFF} as a particular case. It is worth noting that, if we use the
vielbein (\ref{svielb}) in a Dirac operator approach to gravity
 \cite{KW}, the Dirac operator is still self-adjoint and obeys Eq.(8),
provided we take for the $\Gamma^5$ the appropriate ``curved" one.}.

Having a vielbein we can construct a metric tensor $G$. We will think
of $G$ as a functional  \cite{Co2} $ G : \Omega^1 \times \Omega^1
\longrightarrow {\cal A}~,$
such that
\begin{equation}\label{metric1}
G(U F, W H) = F^{\dagger}G(U, W) H ~, ~~~
\forall~~ U,W\in \Omega^1~,~ F, H \in \Omega^0.
\end{equation}

In the $E^A$-basis, the metric is taken to be
\begin{eqnarray}\label{fmetric}
&& G(E^A, E^B) = \eta^{AB}~, \nonumber \\
&& ~~~~~~~~~~\eta^{AB} = diag(-1,1,1,1,1)~.
\end{eqnarray}

In the $DX^M$-basis we will have
\begin{eqnarray}\label{cmetric}
G^{MN}=G(DX^M, DX^N) = E^{M~\dagger}_{~~A}\eta^{AB} E^N_{~~B}~,
\end{eqnarray}
where $E^M_{~~A}$ are the inverses of  $E^A_{~~M}$.

It is worth noting that this metric is symmetric with the
particular vielbein (\ref{svielb}), but is not in the general case allowed by
$Z_2$-noncommutative geometry.

Following Connes \cite{Co2}, we define the connection through a covariant
derivative
$\nabla$
\begin{eqnarray}\label{conn}
&& \nabla : \Omega^1\longrightarrow  \Omega^1
\otimes_{{\cal A}} \Omega^1~, \nonumber \\
&& \nabla(U F) = (\nabla U) F + U \otimes D F~.
\end{eqnarray}
Here the tensor product
$\Omega^1 \otimes_{{\cal A}} \Omega^1$ is generated by
elements $\{ U_1 \otimes U_2 ; U_1, U_2 \in \Omega^1\}$ with the
relation $U_1 F\otimes U_2 = U_1 \otimes F U_2$
for any $F \in \Omega^0$.

A connection is equivalently given by a set of connection one-forms
$\Omega^A_{~~B} \in \Omega^1$, the relation being
\begin{equation}\label{conn1}
\nabla E^A = E^B \otimes \Omega^A_{~~B}~.
\end{equation}

The Cartan structure equations define torsion and curvature of a
given connection as follows
\begin{eqnarray}
T^A = D E^A - E^B \wedge \Omega^A_{~~B}~, \label{tors} \\
R^A_{~~B} = D \Omega^A_{~~B} + \Omega^A_{~~C}
\wedge \Omega^C_{~~B}~, \label{curv}
\end{eqnarray}
where $T^A$ and $R^A_{~~B}$ are $2$-forms.

As in the case of ordinary Riemannian geometry, we can impose the
torsion free condition $T^A = 0$~. Then the structure equation
(\ref{tors}) reduces to
\begin{equation}
D E^A = E^B \wedge \Omega^A_{~~B}~, \label{0tors}
\end{equation}

The connection is said to be a Levi-Civita one it it is also metric compatible,
that is if  it satisfies
\begin{equation}\label{comp}
D(G(U, W)) = \widetilde{G} (\nabla U, W) +
\widetilde{G} (U, \nabla W)~, ~\forall ~ U, W \in \Omega^1~.
\end{equation}
Here $\widetilde{G}$ is the extension of the metric given by
\begin{eqnarray}\label{extmetr}
&& \widetilde{G}(U_1 \otimes U_2, W) = U_2^{\dagger}
G(U_1, W)~, \nonumber \\
&& \widetilde{G}(U, W_1 \otimes W_2) = G(U, W_1)
W_2~,~~\forall~~ U, U_1, U_2~
W, W_1, W_2~ \in \Omega^1~,
\end{eqnarray}
where $^{\dagger}$ denotes the adjoint.

By using the fact that we have a self-adjoint vielbein, condition
(\ref{comp}) gives
\begin{equation}\label{comp1}
\Omega^A_{~~C} \eta^{CB} + \eta^{AC} \Omega^B_{~~C} = 0~.
\end{equation}

The structure equation (\ref{0tors}) and condition (\ref{comp1})
determine
the connection $1$-forms $\Omega^A_{~~B}$ uniquely. From
$\Omega^A_{~~B}$ we can
derive the curvature $2$-forms $R^A_{~~B}$ and their components
$R^A_{~~BCD}$, $R^A_{~~B} = E^C \wedge E^D R^A_{~~BCD}$.

After a little algebra, the scalar curvature
$R = R^A_{~~BAD} \eta^{BD}$ is found to be
\begin{equation}\label{scurv}
R = R_4 - 2 \frac{\Box \phi} {\phi} - \frac{1}{4}
\Psi_{ab} \Psi^{ab}~,
\end{equation}
where
\begin{eqnarray}
&& R_4 ~~~~{\rm is~ the~ 4-dimensional~ Ricci~ scalar} \nonumber \\
&& \Box = \nabla_\mu \partial_{\mu}~, ~~\nabla_{\mu}~~
{\rm is~ the~
4-dimensional~ covariant~ derivative} \nonumber \\
&& \Psi_{ab} = e^{\mu}_a  e^{\nu}_b [ ( \partial_{\mu} a_\nu -
\partial_{\nu} a_\mu ) + a_\mu \frac{\partial_\nu \phi}{\phi} -
a_\nu \frac{\partial_\mu \phi}{\phi} ]~.
\end{eqnarray}

We can redifine the vector field $a_\mu \rightarrow \phi a_\mu$ and
obtain
\begin{equation}\label{FINAL}
R=R_4 -{2\Box\phi\over \phi} -{1\over 4} \phi^2 f_{\mu\nu}f^{\mu\nu}~,
\end{equation}
where $ f_{\mu\nu} = \partial_\mu a_\nu -  \partial_\nu a_\mu$.

The expression for the scalar curvature $R$ in (\ref{FINAL}) is identical to
that in Kaluza-Klein theory  \cite{KK}, when one retains only the zero-modes in
the expansion with respect to the fifth coordinate  and assumes that the fields
are independent of this coordinate. We could then obtain the action by a
suitable integration of $\sqrt{-det|G|} R$. Here $det|G|$ denotes the
determinant of our generalized metric defined above and is given as follows
\begin{eqnarray}
det|G|& \doteq &{1\over 5!}{\epsilon }_{N_1 N_2 N_3 N_4 N_5}
{\epsilon}_{M_1 M_2 M_3 M_4 M_5} G^{N_1 M_1} G^{N_2 M_2} G^{N_3 M_3}
G^{N_4 M_4} G^{N_5 M_5} \nonumber\\
      & = &
{1\over 4!}{\epsilon}_{ \nu_1 \nu_2 \nu_3 \nu_4}
{\epsilon}_{\mu_1
\mu_2 \mu_3 \mu_4 } G^{\nu_1\mu_1} G^{\nu_2 \mu_2} G^{\nu_3 \mu_3} G^{\nu_4
\mu_4} G^{55} \equiv  det|g|\phi~{\bf 1}
\end{eqnarray}
where $det|g|$ is the determinant of the 4-dimensional metric and
${\epsilon}$
are fully antisymmetric Levi-Civita tensors.
If we take the trace of the
$2\times 2 $ matrix as the integration over the discrete coordinate,
we obtain
\begin{equation}
S\sim \int d^4x \sqrt{-det|g|} \phi R,
\end{equation}
which reproduces the Kaluza-Klein action up to a proportional
constant. The
parameter $m$ replaces the radius of the
compactified circle in the fifth dimension of the Kaluza-Klein
theory\footnote{Our action differs from
the action obtained previously  \cite{CFF,Si,KW} by a factor of $\phi$
even if we set
$a=0$ to have the same vielbein.}. Thus, in
the model based on the vielbein (\ref{svielb}), we have the dimensionally
reduced Kaluza-Klein theory with massless tensor, vector and scalar fields. The
most general vielbein of Eq.(\ref{gvielb}) yields a full Kaluza-Klein theory,
with a finite number of fields.
The field content of the full theory contains a pair of tensors,
a pair of vectors
and a pair of scalar fields without imposing any truncation condition.
In each pair, one field is massless and the other
is its massive excitation. The inconsistencies of the Kaluza-Klein theory that
arise when one truncates the spectrum by including only a finite number of
massive modes are absent in the noncommutative geometric
approach.
These matters and further details of mathematical formalism
will be presented elsewhere  \cite{LVW}.

\bigskip

\bigskip

\noindent
{\bf Acknowledgments.}

This work was supported in part by the U.S. Department of Energy under contract
number DE-FG02-85ER40231.

The work of G.L. is partially supported by the Italian `Ministero
dell' Universit\`a e della Ricerca Scientifica'.

G.L. is grateful to Prof. A.P. Balachandran for his kind hospitality
in Syracuse.
He also wishes to thank
him, L. Chandar, E. Ercolessi, A. Momen and P. Teotonio for many
useful discussions on noncommutative geometry.

N.A.V. is grateful to Professor A.Zichichi for a World Laboratory
scholarship.

\end{document}